\documentclass[epj]{svjour}
\usepackage{graphicx}

\ifnum\lefthyphenmin<2\lefthyphenmin=2\fi
\ifnum\righthyphenmin<2\righthyphenmin=2\fi

\def\letterone{x}
\def\lettertwo{y}
\def\width{W}
\def\difference{\Delta}
\def\deltal{\delta}
\def\statevec{\psi}

\begin{document}

\title{High precision simulations of the longest common subsequence problem}
\author{R.~Bundschuh}
\institute{Department of Physics, University of California at San Diego,
La Jolla, CA  92093-0319, U.S.A, \email{rbund@matisse.ucsd.edu}}

\date{11 April 2001}

\abstract{
The longest common subsequence problem is a long studied
prototype of pattern matching problems. In spite of the effort
dedicated to it, the numerical value of its central quantity, the
Chv\'atal-Sankoff constant, is not yet known. Numerical estimations of
this constant are very difficult due to finite size effects. We
propose a numerical method to estimate the Chv\'atal-Sankoff constant
which combines the advantages of an analytically known functional form
of the finite size effects with an efficient multi-spin coding
scheme. This method yields very high precision estimates of the
Chv\'atal-Sankoff constant. Our results correct earlier estimates for
small alphabet size while they are consistent with (albeit more
precise than) earlier results for larger alphabet size.}


\maketitle

\section{Introduction}

Pattern matching problems have been of central interest in such
different areas of science as image processing, speech recognition,
time series analysis, and biological sequence comparison for a large
number of years. In many practical applications of pattern matching,
it is important to be able to characterize the amount and/or strength
of patterns found in {\em random data} in order to be able to
distinguish meaningful patterns from random ones~\cite{karl90}. This
requires a {\em quantitative theory} of pattern matching procedures
which does not only characterize universal characteristics like
scaling laws but which also provides specific numerical values for
non-universal quantities.

One of the simplest pattern matching problems is the longest common
subsequence (LCS) problem. It is a special case of the sequence
comparison problem, which is a very important tool in modern molecular
biology~\cite{wate94}. As a prototype for pattern matching it has
enjoyed the attention of mathematicians and computer scientists for a
long time~\cite{chva75,hirs78,deke79,stee82,chva83,deke83,apos87,%
danc94a,danc94b,demo99,demo00}.  One of the central quantities in the
LCS problem is the average fraction of matches in the longest common
subsequence of two very long random strings also known as the {\em
Chv\'atal-Sankoff constant} $a_c$~\cite{chva75}. However, in spite of
the relative simplicity of the LCS problem and the long time it has
been studied, the numerical value of the Chv\'atal-Sankoff constant is
still unknown. There has been a long standing conjecture
\begin{equation}\label{eq_conjecture}
a_c\stackrel{?}{=}\frac{2}{\sqrt{c}+1}.
\end{equation}
by Arratia~\cite{stee97} for its value based on numerical simulations
where $c$ is the size of the alphabet from which the sequences are
chosen at random. This conjecture has been recently shown to hold for
a {\em first-passage percolation version} of the LCS problem using the
analogy~\cite{hwa96} to surface growth as described by the
Kardar-Parisi-Zhang (KPZ~\cite{kard86}) equation~\cite{bund00} and a
cavity method~\cite{demo99,demo00}. In the language of surface growth
the Chv\'atal-Sankoff constant takes the interpretation of the average
growth velocity of the surface or of the ensemble averaged free energy
per length in the directed polymer picture which is equivalent to the
KPZ equation.

For the original Chv\'atal-Sankoff constant (i.e., not the
first-passage percolation version) a variety of {\em upper and lower
bounds} have been rigorously proven in the mathematical
literature~\cite{chva75,hirs78,chva83,deke83,danc94a,danc94b} which
are all consistent with Arratia's conjecture. Short of a rigorous
result for the value of the Chv\'atal-Sankoff constant, {\em extensive
numerical studies} are the only way to shed some light on the validity
of Arratia's conjecture. The difficulty with numerical studies is that
the Chv\'atal-Sankoff constant is defined as an asymptotic quantity in
the limit that the sequences the longest common subsequence of which
is computed are infinite. Simulations at finite sequence length always
observe {\em finite size effects} the size and even functional form of
which is not known. In fact, two recent numerical
studies~\cite{danc94b,demo99,demo00} assume different functional forms
of the finite size effect. Thus, they come to {\em different
conclusions} about the value of the Chv\'atal-Sankoff constant
although both studies agree on the fact that the Chv\'atal-Sankoff
constant {\em differs} from its conjectured value
Eq.~(\ref{eq_conjecture}).

In this communication, we will remedy this dilemma in two different
ways. First, we will use a slightly different geometry in which we
perform our simulations.  The advantage of this geometry is that the
form of the finite size behavior is {\em theoretically known} from the
analogy to surface growth~\cite{hwa96} and does not have to be
inferred from the numerical data. We will prove this finite size
behavior for the first-passage percolation case explicitly. While this
result reproduces the known value of the first-passage percolation
version of the Chv\'atal-Sankoff constant, it additionally provides
the {\em complete finite-size correction} in terms of Legendre
polynomials and an {\em explicit formula} for the prefactor of the
leading finite size correction term. We will use this to validate our
numerical procedure. Additionally, we will reformulate the LCS problem
in a way that allows for a {\em multi-spin coding scheme}, in which
$64$ lattice sites are updated in parallel by simple bitwise logical
operations. This extremely efficient code allows us to average over a
large number of systems with several million letters each. This yields
the value of the Chv\'atal-Sankoff constant with a precision of one
part in $10^{6}$.

\section{Longest Common Subsequence problem}

The LCS problem is defined as follows: Given two sequences
$\vec\letterone=\letterone_1\ldots \letterone_N$ and
$\vec\lettertwo=\lettertwo_1\ldots \lettertwo_N$ of letters
$\letterone_i$ and $\lettertwo_j$ independently chosen with equal
probabilities from an alphabet of size $c$, find their longest common
subsequence. A common subsequence of length $\ell$ is given by a set
of $(i_k,j_k)$'s with $1\le i_1<\ldots< i_\ell\le N$ and $1\le
j_1<\ldots<j_\ell\le N$ such that $\letterone_{i_k}=\lettertwo_{j_k}$
for all $k\in\{1,\ldots,\ell\}$. We denote the length of the longest
common subsequence of two given sequences by $L(N)$. It is a random
variable since it depends on the specific choice of the sequences.
Its expectation value averaged over the ensemble of all sequence pairs
is $\langle L(N)\rangle$. Chv\'atal and Sankoff noted~\cite{chva75}
that this expectation value is asymptotically proportional to the
length of the sequences, i.e., $\langle L(N)\rangle\approx a_c N$. The
prefactor $a_c$ is now called the Chv\'atal-Sankoff constant and
depends only on the size $c$ of the alphabet. To find high precision
estimates for this value for different alphabet sizes $c$ is the
purpose of this publication.

\begin{figure}
\begin{center}
\includegraphics[width=\columnwidth]{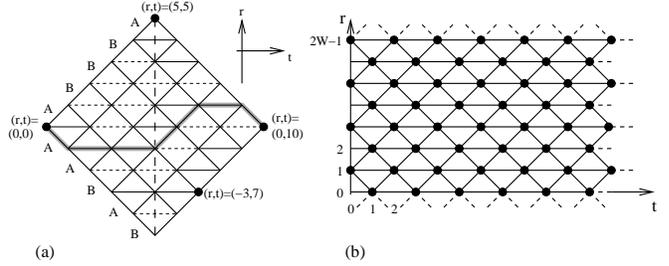}
\caption{Lattice representation of the longest common subsequence
problem. (a) shows how the two sequences \protect$ABBBA$ and
\protect$AABAB$ are written at the edges of the lattice. The
horizontal bonds of the lattice correspond to matches with a weight
one (solid lines) or to mismatches with a weight zero (dashed lines.)
The diagonal bonds correspond to omitted letters. Each common
subsequence corresponds to a path from the leftmost to the rightmost
point of the lattice. The length of the common subsequence is the sum
over all weights of the bonds which the path passed. The path shown as
an example has a weight of three and corresponds to the maximal common
subsequence $ABA$. The figure also shows how we use the coordinates
\protect$r$ and \protect$t$ in order to denote lattice points and how
the diamond shaped lattice can be split into two triangular lattices
(long dashed line.) (b) shows the rectangular lattice which we use
instead of the lattice of (a) in order to be able to handle the
effects of the finite lattice size.}
\label{fig_alignl}
\end{center}
\end{figure}

Numerically, the length of the longest common subsequence of two given
sequences can be calculated very efficiently using a transfer matrix
algorithm. This transfer matrix algorithm calculates the auxiliary
quantity $\ell_{i,j}$ which is defined to be the length of the longest
common subsequence of the substrings $\letterone_1\ldots\letterone_i$
and $\lettertwo_1\ldots\lettertwo_j$.  The recursion for $\ell_{i,j}$
is
\begin{equation}
\ell_{i,j}=\max\{\ell_{i-1,j-1}+\eta_{i,j},\ell_{i-1,j},\ell_{i,j-1}\}
\end{equation}
where
\begin{equation}
\eta_{i,j}=\left\{\begin{array}{ll}1&\mbox{if }\letterone_i=\lettertwo_j\\
0&\mbox{otherwise}\end{array}\right..
\end{equation}
This recursion is completed by the initial conditions
\begin{equation}\label{eq_initial}
\ell_{i,0}=\ell_{0,j}=0.
\end{equation}
The length of the longest common subsequence of the original strings
$\vec\letterone$ and $\vec\lettertwo$ is then $L(N)=\ell_{N,N}$.  The
recursion equation can be visualized by assigning the $\ell_{i,j}$ to
the nodes of the lattice shown in
Fig.~\ref{fig_alignl}(a)~\cite{need70,alex94}. The horizontal bonds
then correspond to the $\eta_{i,j}$ and the diagonal bonds correspond
to omitting a letter from one of the sequences. The recursive
algorithm calculates the path from the origin to the end point of this
lattice with the largest weight as given by the sum over the bond
weights through which the path passes. This is clearly a discrete
model of a directed polymer (the path) in a random medium~\cite{hwa96}
(the random matches and mismatches $\eta_{i,j}$.)  To make this more
explicit, we will use ``spatial'' and ``temporal'' coordinates $r$ and
$t$ instead of the indices $i$ and $j$ to refer to lattice points as
shown in Fig.~\ref{fig_alignl}(a). In this coordinate system the
dynamics given by the recursive algorithm reads
\begin{equation}\label{eq_recurs}
\ell(r,t+1)=\max\left\{\begin{array}{l}\ell(r,t-1)+\eta(r,t)\\
\ell(r-1,t)\\\ell(r+1,t)\end{array}\right\}.
\end{equation}

The total length $L(N)=\ell(0,2N)$ of the longest common subsequence
corresponds to the total free energy of the directed polymer. This is
clearly an extensive quantity and the proportionality constant, the
free energy per length, is the Chv\'atal-Sankoff constant. It can in
principle be calculated numerically by running the
algorithm~(\ref{eq_recurs}) for a large number of sequences and
averaging the free energies obtained. As already mentioned the
difficulty with this procedure is that the free energy has terms which
are subleading in the length $N$ of the sequences. Thus, the ratio
$\langle L(N)\rangle/N$ only converges towards the Chv\'atal-Sankoff
constant $a_c$ as the sequences become very long. For high precision
measurements, the rate of this convergence has to be taken into
account for a proper extrapolation towards infinite sequence
lengths. This is relatively hard for the diamond shaped lattice, since
there is no theoretical understanding of its finite size effects.  The
only reliable result is a {\em bound}~\cite{alex94} $\langle
L(N)\rangle/N\le a_c+O({N^{-1/2}}\log N)$ on the finite size
dependence. Dan{\v c}{\'\i}k~\cite{danc94b} assumed that this bound
represents the exact functional form of the finite size correction and
used it to fit his numerical data generated from four pairs of
sequences of length up to one million to arrive at a value of
$a_{c=2}=0.81225\pm0.00025$ for a two letter alphabet. More recently,
Boutet de Monvel tried to extract the finite size behavior directly
from the numerical data without any theoretical foundation and
concluded $\langle L(N)\rangle/N\approx a_c+O({N^{-1/2}}/\log N)$
which lead him to a value\footnote{Boutet de Monvel does not give an
actual statistical error of his estimates. Instead he gives three
different estimates for the Chv\'atal-Sankoff constant at each
alphabet size. These three values stem from three different ways to
treat the finite size effects. The statistical error cited here
reflects the spread of these three values.} of
$a_{c=2}=0.81233\pm0.00005$~\cite{demo99,demo00}.

\section{LCS on a rectangular lattice}\label{sec_rectangularlcs}

In light of the confusion on the handling of finite size effects,
we will consider a rectangular lattice of a given width $\width$ as
the one shown in Fig.~\ref{fig_alignl}(b). In order to minimize the
finite size effects, we apply periodic boundary conditions in the
spatial (vertical) direction. For any fixed $\width$ the dynamics
given by Eq.~(\ref{eq_recurs}) will after some startup phase become
stationary which allows us to measure
\begin{eqnarray}\label{eq_a0wdef}
a_c(\width)&\equiv&
\lim_{t\to\infty}\frac{1}{t}\langle\ell(\width-1,2t)\rangle\\\nonumber
&=&\lim_{t\to\infty}\frac{1}{t}\frac{1}{2\width}
\sum_{k=0}^{\width-1}\langle\ell(2k+1,2t)+\ell(2k,2t-1)\rangle
\end{eqnarray}
with nearly arbitrary precision if we use long enough sequences (note
also, that in this geometry we have to perform only $\width N$ instead
of $N^2$ elementary steps as the one given in Eq.~(\ref{eq_recurs}) on
the diamond shaped lattice which allows for much larger sequence
lengths $N$.) This geometry directly corresponds to the finite size
geometry usually imposed in studies of directed polymers in a random
medium and equivalent KPZ surface growth problems~\cite{krug91a}. From
these studies it is well known~\cite{krug90}, that
\begin{equation}\label{eq_longtimefsc}
a_c(\width)=a_c(\infty)-\frac{b_c}{\width}+\mbox{higher order terms}
\end{equation}
with some constant $b_c$. We argue in App.~\ref{app_changelattice} that
\begin{equation}
a_c(\infty)\equiv\lim_{\width\to\infty}a_c(\width)=a_c,
\end{equation}
i.e., that calculating $a_c(W)$ for fixed $W$ and taking the limit
$W\to\infty$ actually yields the Chv\'atal-Sankoff constant $a_c$. The
advantage of this procedure over other numerical approaches is that we
know the functional form of the finite size dependence of $a_c(W)$ in
advance and can extract the Chv\'atal-Sankoff constant $a_c$ by fitting
numerically estimated $a_c(W)$ to the functional form
\begin{equation}\label{eq_finitesize}
a_c(\width)\approx a_c-\frac{b_c}{\width}.
\end{equation}

At this point we want to remark for the reader familiar with the
surface growth analogy of the LCS problem that in the surface growth
language there are {\em two} finite size correction formulae
known~\cite{krug90}. If the surface grows for a very long time $t$ but
is restricted to a width $\width\ll t^{2/3}$
Eq.~(\ref{eq_longtimefsc}) applies. In the opposite regime, i.e., for
$\width\gg t^{2/3}$, there is a different scaling behavior which now
depends on $t$ instead of $\width$. It appears that the diamond shaped
lattice of the original LCS problem is in this transient phase since
we have $t=2\width$. However, the known finite size correction formula
in this regime applies only if the surface grows from a {\em flat}
substrate while the diamond lattice forms a {\em wedged} substrate
leading to the more difficult finite size effects observed numerically
by Boutet de Monvel~\cite{demo99,demo00}. It turns out, though, that
the flat substrate can also be translated into the LCS language: It
requires cutting the diamond shaped lattice of size $t$ into two
triangles as indicated by the long dashed line in
Fig.~\ref{fig_alignl}(a), applying the recursion
equation~(\ref{eq_recurs}) with the diamond boundary conditions
Eq.~(\ref{eq_initial}) within the left triangle, and then calculating
\begin{equation}
\widetilde{a}_c(t)\equiv\frac{2}{t}
\langle\max_{r\in\{-t,-t+2,\ldots t-2,t\}}\ell(r,t)\rangle.
\end{equation}
It can be shown~\cite{alex94} that $\lim_{t\to\infty}\widetilde{a}_c(t)=a_c$
and the finite size correction should for this geometry be~\cite{krug90}
\begin{equation}
\widetilde{a}_c(t)=a_c-\frac{\widetilde{b}_c}{t^{2/3}}
+\mbox{higher order terms}.
\end{equation}
This explicit finite size dependence formula could also be used in
order to get high precision estimates of the Chv\'atal-Sankoff
constant.  However, as we will see below the rectangular geometry is
advantageous in terms of the numerical evaluation of the recursion
equation~(\ref{eq_recurs}). Thus, we will use the rectangular
geometry in our current study.

For a better theoretical understanding of the longest common
subsequence problem on the rectangular lattice and in order to
formulate a very efficient algorithm for its simulation it is helpful
to concentrate on the differences between neighboring $\ell(r,t)$. In
order to preserve the symmetry of the lattice shown in
Fig.~\ref{fig_alignl}(b) we define them as
\begin{equation}
\difference(r,t)\equiv\left\{\begin{array}{ll}
\ell(r,t+1)-\ell(r+1,t)&\mbox{for $r+t$ even}\\
\ell(r+1,t+1)-\ell(r,t)&\mbox{for $r+t$ odd}\end{array}\right.
\end{equation}
From the recursion equation~(\ref{eq_recurs}) it can be easily checked
by induction that $\difference(r,t)\in\{0,1\}$. They can thus be
considered {\em Boolean variables}. In terms of these quantities, the
recursion equation~(\ref{eq_recurs}) can be rewritten as
\begin{eqnarray}\label{eq_ldiffrecurs1}
\difference(r,t)\!&\!=\!&
\!\max\!\!\left\{\begin{array}{l}\eta(r,t)-\difference(r,t-1)\\
\difference(r-1,t-1)-\difference(r,t-1)\\0\end{array}\!\!\right\}
\\\label{eq_ldiffrecurs2}
\difference(r\!-\!1,t)\!&\!=\!&
\!\max\!\!\left\{\begin{array}{l}\eta(r,t)-\difference(r-1,t-1)\\
\difference(r,t-1)-\difference(r-1,t-1)\\0\end{array}\!\!\right\}
\end{eqnarray}
for even $r+t$. Note, that this formulation does not make any
reference to the absolute lengths $\ell(r,t)$ of the longest common
subsequences.  The dynamics of the LCS problem is thus completely
captured by the dynamics of the length differences as given by
Eqs.~(\ref{eq_ldiffrecurs1}) and~(\ref{eq_ldiffrecurs2}). Every
configuration of the $\ell(r,t)$ can be equally well represented by
a configuration of the $\difference(r,t)$. Reversely, a configuration
of the $\difference(r,t)$ represents a configuration of the
$\ell(r,t)$, if the consistency condition
\begin{equation}\label{eq_consistcond}
\sum_{r=0}^{2\width-1}(-1)^r\difference(r,t)=0
\end{equation}
for the validity of the periodic boundary conditions in $\ell(r,t)$ is
fulfilled. The alternating sum on the left hand side of
Eq.~(\ref{eq_consistcond}) is conserved under the dynamics given by
Eq.~(\ref{eq_ldiffrecurs1}) and~(\ref{eq_ldiffrecurs2}) and is thus
always fulfilled, if it is fulfilled for the initial choice of the
$\difference(r,t=0)$.

Slaved to the dynamics of the $\difference$ is the evolution of
the {\em average length}
\begin{displaymath}
\overline{\ell}(t)\equiv\left\{\begin{array}{ll}
\displaystyle\frac{1}{2\width}\sum_{k=0}^{\width-1}[\ell(2k,t\!-\!1)
+\ell(2k\!+\!1,t)]
&\mbox{for $t$ even}\\[5pt]
\displaystyle\frac{1}{2\width}\sum_{k=0}^{\width-1}[\ell(2k,t)
+\ell(2k\!+\!1,t\!-\!1)]
&\mbox{for  $t$ odd}
\end{array}\right.
\end{displaymath}
of the longest common subsequence which changes by
\begin{equation}\label{eq_defaveragel}
\overline{\ell}(t+1)-\overline{\ell}(t)=\frac{1}{2\width}\sum_{k=0}^{\width-1}
\left\{\begin{array}{ll}\deltal(2k,t)&\mbox{for $t$ even}\\
\deltal(2k+1,t)&\mbox{for $t$ odd}\end{array}\right.
\end{equation}
with
\begin{equation}
\deltal(r,t)\!\equiv\!\ell(r,t\!+\!1)\!-\!\ell(r,t\!-\!1)\!=\!
\max\left\{\!\begin{array}{l}\eta(r,t)\\\difference(r,t\!-\!1)\\
\difference(r\!-\!1,t\!-\!1)\end{array}\!\right\}
\end{equation}
in every time step. Together with Eq.~(\ref{eq_a0wdef}) it thus yields
\begin{equation}\label{eq_localadvance}
a_c(\width)=\langle\deltal(r,t)\rangle.
\end{equation}

The rewriting of $a_c(W)$ in terms of $\langle\deltal\rangle$
opens the possibility of a very fast implementation of the longest
common subsequence problem. We note that Eqs.~(\ref{eq_ldiffrecurs1})
and~(\ref{eq_ldiffrecurs2}) transform a pair of length differences
into another pair of length differences independent of the other pairs
at the same time $t$. According to the lattice structure shown in
Fig.~\ref{fig_alignl}(b) this happens for different ways of pairing at
different times. If we interpret the $\difference(r,t)$ and
$\eta(r,t)$ as Boolean variables, we can give
Eqs.~(\ref{eq_ldiffrecurs1}) and~(\ref{eq_ldiffrecurs2}) the form
\begin{eqnarray}\label{eq_ldiffboolean1}
\difference(r,t)\!&\!=\!&\![\neg\,\difference(r,t\!-\!1)]\wedge
[\eta(r,t)\!\vee\!\difference(r\!-\!1,t\!-\!1)]\\\label{eq_ldiffboolean2}
\hspace*{-5mm}
\difference(r\!-\!1,t)\!&\!=\!&\!
[\neg\,\difference(r\!-\!1,t\!-\!1)]\wedge
[\eta(r,t)\!\!\vee\difference(r,t\!-\!1)]
\end{eqnarray}
Thus, we store $32$ of the $\difference(r,t)$ for even $r$ and the
corresponding $32$ of the $\difference(r,t)$ for odd $r$
simultaneously as the single bits of one integer variable each. $32$ of
the match-mismatch variables $\eta(r,t)$ can be stored in the same
way. Then, Eqs.~(\ref{eq_ldiffboolean1}) and~(\ref{eq_ldiffboolean2})
can be performed in parallel on these $64$ lattice sites by bitwise
logical operations. After these operations the odd sites are shifted
by one bit to the right so that they pair up with their new even site
partners and the same procedure is repeated.  After shifting the odd
sites one bit back to the left the algorithm can proceed from the
beginning

In order to measure $a_c(\width)$ we note that the increase
$\deltal(r,t)$ in Eq.~(\ref{eq_defaveragel}) is given by the Boolean
equation
\begin{equation}\label{eq_advance}
\deltal(r,t)=
\eta(r,t)\vee\difference(r,t\!-\!1)\vee\difference(r\!-\!1,t\!-\!1)
\end{equation}
and can thus also be easily calculated for $32$ pairs of sites with
two bitwise or operations. A bit-count then gives the value of $a_c(\width)$
after averaging over many time steps and configurations of the disorder
$\eta(r,t)$.

\section{First passage percolation}\label{sec_firstpassage}

We verify the algorithm and the finite size behavior given by
Eq.~(\ref{eq_finitesize}) on the first passage percolation version of
the LCS problem. For this system we are not only able to prove
explicitly the scaling form Eq.~(\ref{eq_finitesize}) but we can
calculate the full finite size correction and specifically the
prefactor $\widehat{b}_c$ of its leading term. The first passage
percolation version of the LCS problem also obeys the dynamics given
by Eqs.~(\ref{eq_ldiffrecurs1}) and~(\ref{eq_ldiffrecurs2}). The only
difference is, that instead of choosing random sequences
$\letterone_1\ldots\letterone_N$ and $\lettertwo_1\ldots\lettertwo_N$
and calculating the local matches $\eta(r,t)$ from these sequences, we
use the variables $\widehat{\eta}(r,t)$ which are taken to be {\em
independent} identically distributed random variables with
\begin{equation}
\widehat{\eta}(r,t)=\left\{\begin{array}{ll}1&\mbox{with probab. $1/c$}\\
0&\mbox{with probab. $1-1/c$}\end{array}\right..
\end{equation}
The analog $\widehat{a}_c$ of the Chv\'atal-Sankoff constant for this
problem is known to be $\widehat{a}_c=2/(\sqrt{c}+1)$ which had been
conjectured by Arratia for the $a_c$ of the LCS problem. This result
follows, e.g., as a consequence of an exact mapping~\cite{bund01}
between the LCS problem and the discrete time asymmetric exclusion
process~\cite{krug91b,derr98a}, a well studied exactly solvable model
of the KPZ surface growth universality class.  (The mapping basically
interprets the length differences $\difference(r,t)$ as the site
occupation numbers of the asymmetric exclusion process.) Below, we
will derive the result $\widehat{a}_c=2/(\sqrt{c}+1)$ and its finite
size correction without explicit reference to this mapping.

Eqs.~(\ref{eq_ldiffboolean1}) and~(\ref{eq_ldiffboolean2}) can be
interpreted in terms of a pair
$(\difference(r\!-\!1,t-1),\difference(r,t-1))\in
\{(0,0),(0,1),(1,0),(1,1)\}$ mapped onto another pair
$(\difference(r-1,t),\difference(r,t))\in\{(0,0),$ $(0,1),(1,0),(1,1)\}$
with probabilities depending on the possible values of
$\widehat{\eta}(r,t)$.  This mapping can be written as the matrix
\begin{equation}
T_0\equiv\left(
\begin{array}{cccc}1\!-\!\frac{1}{c}&0&0&1\\0&0&1&0\\0&1&0&0\\
\frac{1}{c}&0&0&0\end{array}
\right),
\end{equation}
in the basis $|00\rangle$, $|01\rangle$, $|10\rangle$, $|11\rangle$. A
distribution of possible states of the system can be interpreted as a
normalized vector $|\statevec\rangle$ in the $2^{2\width}$ dimensional space
of all possible configurations of the $2\width$ variables
$\difference(r,t)$ at a fixed time. In a time step with even $t$, each
neighboring pair of length differences is transformed according to the
matrix $T_0$. Thus, the whole state distribution $|\statevec\rangle$ is
transformed into the vector $T|\statevec\rangle$ with
\begin{equation}
T\equiv\bigotimes_{i=1}^\width T_0.
\end{equation}
At odd time steps the pairing of the length differences is different.
In order to describe the transformation at odd time steps, we introduce
the translation operator $C$, which shifts all length differences
$\difference(r,t)$ by one spatial position to the right, i.e.,
\begin{equation}
C|\difference_{2\width-1}\ldots\difference_1\difference_0\rangle
\equiv|\difference_0\difference_{2\width-1}\ldots\difference_1\rangle.
\end{equation}
The time evolution at odd times is then given by a shift of all length
differences to the left, application of the even time evolution
operator $T$ and a shift of all length differences back to the right,
i.e., by $CTC^{-1}$.  After two time steps, the state vector
$|\statevec\rangle$ is thus transformed into the state vector
$CTC^{-1}T|\statevec\rangle$. The stationary state is the eigenvector
$|\statevec_0\rangle$ of $CTC^{-\!1}T$ with the eigenvalue one which is
consistent with the condition~(\ref{eq_consistcond}). It is easy to
check, that this eigenvector is given by
\begin{equation}\label{eq_stationarystate}
|\statevec_0\rangle\!\equiv\!\frac{1}{{\cal N}_\width}\hspace*{-3mm}
\sum_{\{\difference_j:
\!\!\sum\limits_{i=0}^{2\width-1}\!\!\!(-\!1)^i\difference_i=0\}}
\hspace*{-3mm}
\left(\!\frac{1}{\sqrt{c}}\!\right)^{\!\!\!\!\sum\limits_{i=0}^{2\width-1}
\!\!\!\!\difference_i}
|\difference_{2\width-1}\ldots \difference_0\rangle.
\end{equation}
with the combinatorial normalization factor
\begin{equation}\label{eq_normfactor}
{\cal N}_\width\!=\!\sum_{\mu=0}^{\width}\!\left(\frac{1}{c}\right)^{-\mu}\!
{\width\choose\mu}^2\!\!
\!=\!\left(1-\frac{1}{c}\right)^\width
\!\!\!P_\width\left(\frac{c+1}{c-1}\right),
\end{equation}
where $P_\width$ is the Legendre polynomial of degree $\width$. Since
$CTC^{-1}T$ restricted to the subspace of state vectors consistent
with condition~(\ref{eq_consistcond}) is irreducible and neither
the eigenvector $|\statevec_0\rangle$ nor $CTC^{-1}T$ have any negative
entries, it is the unique stationary state of the system according to
the Perron Frobenius theorem~\cite{bapa97}.

The analog $\widehat{a}_c(\width)$ of the Chv\'atal-Sankoff constant for
this modified system can be calculated from the knowledge of this
stationary state via Eqs.~(\ref{eq_localadvance}) and~(\ref{eq_advance}).
Since $\widehat{\eta}(r,t)$ is uncorrelated from $\difference(r-1,t-1)$
and $\difference(r,t-1)$ we get
\begin{eqnarray}\label{eq_calcahat}
\widehat{a}_c(\width)\!&=&\!
\langle\widehat{\eta}(r,t)\vee\difference(r,t-1)\vee
\difference(r-1,t-1)\rangle\\\nonumber
&=&\!1\!-\!\left(1\!-\!\frac{1}{c}\right)
\langle[\neg\,\difference(r,t\!-\!1)]\wedge
[\neg\,\difference(r\!-\!1,t\!-\!1)]\rangle\\\nonumber
&=&\!1\!-\!\left(1\!-\!\frac{1}{c}\right)
\Pr\{\difference(r,t\!-\!1)\!=\!\difference(r\!-\!1,t\!-\!1)\!=\!0\}.
\end{eqnarray}
The probability that two neighboring length differences vanish
simultaneously can be directly calculated from
the explicitly known stationary state Eq.~(\ref{eq_stationarystate}).
Since fixing a pair of the length differences to zero still leaves all
the possibilities to the $\width-1$ remaining pairs, it is
\begin{equation}
\Pr\{\difference(r,t-1)=\difference(r-1,t-1)=0\}=
\frac{{\cal N}_{\width-1}}{{\cal N}_\width}.
\end{equation}
Inserting this into Eq.~(\ref{eq_calcahat}) yields the {\em full finite
size dependence} of the first passage percolation version of the LCS
problem in terms of Legendre polynomials. Expanding these Legendre
polynomials for large degrees $\width$ finally yields the first order
terms
\begin{equation}
\widehat{a}_c(\width)
=\frac{2}{\sqrt{c}\!+\!1}
-\frac{\sqrt{c}\!-\!1}{2(\sqrt{c}\!+\!1)}\frac{1}{\width}
+O\left(\frac{1}{\width^2}\right)
\end{equation}
of this exact finite size correction.

This can be directly compared to the numerical results obtained by the
multi-spin coding algorithm described above. To this end we choose
$1,000$ random initial conditions of the $\difference(r,t=0)$ (that
fulfil Eq.~(\ref{eq_consistcond})) and the same number of random
configurations of the $\widehat{\eta}(r,t)$.  Then, we proceed as
described at the end of Sec.~\ref{sec_rectangularlcs}, i.e., we apply
the recursion equations~(\ref{eq_ldiffboolean1})
and~(\ref{eq_ldiffboolean2}) a large number of times, evaluate
Eq.~(\ref{eq_advance}) in every time step, and generate estimates of
$\widehat{a}_c(\width)$ according to Eq.~(\ref{eq_localadvance}) by
averaging over many time steps and the different configurations of the
disorder.

An important question is how close these estimates of
$\widehat{a}_c(\width)$ come to its true value.  In answering this
question, we have to consider {\em statistical} and {\em systematic}
deviations. The statistical deviations can be measured by looking at
the sample-to-sample fluctuations between the different configurations
of the disorder $\widehat{\eta}(r,t)$ and made as small as we wish by
increasing the number of disorder configurations and the number of
recursion steps we average over. The systematic deviations have to be
studied more carefully. They arise because the Markov dynamics of the
differences $\difference(r,t)$ reaches its stationary state only in
the limit of infinitely many recursion steps if we start from a random
initial state. For a finite number $t_1$ of recursions steps there is
a systematic correction that decays exponentially in $t_1$ with some
decay ``time'' $t_0$.

In order to verify if this exponential approach to the stationary
state plays a role in our numerics, we iterate
Eqs.~(\ref{eq_ldiffboolean1}) and~(\ref{eq_ldiffboolean2}) $10^7$
times for each configuration of the disorder $\widehat{\eta}(r,t)$ and
calculate $10$ different estimates of $\widehat{a}_c(\width)$ by
averaging in Eq.~(\ref{eq_localadvance}) over $t\in]t_1,t_1+10^6]$ for
$t_1\in\{0,10^6,2\cdot10^6,\ldots,9\cdot10^6\}$. If there is a
systematic deviation this should manifest itself in a systematic
dependence of these estimates on the number $t_1$ of recursion steps
performed before the averaging starts.

\begin{figure}
\begin{center}
\includegraphics[width=\columnwidth]{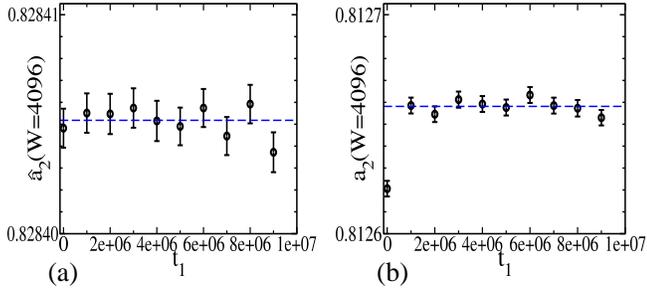}
\caption{Dependence of estimates of the Chv\'atal-Sankoff constant
on the number \protect$t_1$ of recursion steps performed before the
averaging starts. (a) shows estimates of
\protect$\widehat{a}_2(\width)$ for completely uncorrelated disorder,
i.e., for the first-passage percolation problem. They do coincide
within their statistical fluctuations for all \protect$t_1$. The
dashed line shows the average over all points but the first. (b) shows
estimates of \protect$a_2(\width)$ for the full LCS problem including
the disorder correlations. They have a larger statistical fluctuation
than in the first passage percolation case and the first point
deviates significantly from the others. Both effects stem from the
much slower approach to the stationary state in the presence of the
correlations. However, there is no apparent dependence on
\protect$t_1$ for \protect$t_1\ge10^6$ in this case either.  The
dashed line shows the average over the last eight data points.}
\label{fig_decorrel}
\end{center}
\end{figure}

Since the approach to the stationary state has to become the slower
the wider the lattice we have to study this effect for the largest
width $\width\!\!=\!4096$ we plan on using in our numerics.
Fig.~\ref{fig_decorrel}(a) shows the ten estimates of
$\widehat{a}_2(\width=4096)$ which are obtained as described above.
Clearly, there is no noticeable systematic dependence on the number
$t_1$ of recursion steps. All fluctuations of the individual points
are within the statistical uncertainties of each other. While the
statistical uncertainties of the individual points are already on the
order of $\pm10^{-6}$ we can further improve them by averaging over all
the points. To be absolutely certain that the approach to the
stationary state does not influence our result, we take this average
only over the last nine points, i.e., we average over all
$t\in]10^6,10^7]$. This average is indicated as the dashed line in
Fig.~\ref{fig_decorrel}(a).

\begin{figure}
\begin{center}
\includegraphics[width=\columnwidth]{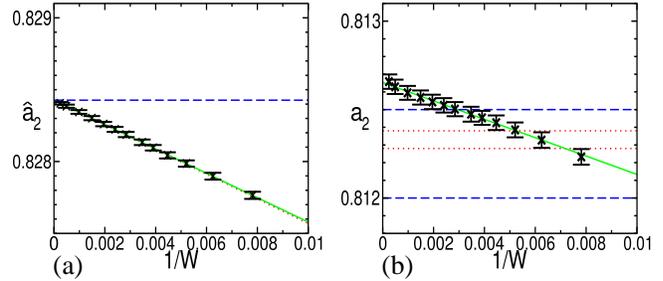}
\caption{Numerical estimates of the Chv\'atal-Sankoff constant for a
two letter alphabet as a function of \protect$1/\width$. The solid
lines are in both cases best fits to a linear function. (a) shows
values \protect$\widehat{a}_2(\width)$ for completely uncorrelated
disorder, i.e., for the first-passage percolation problem. The dashed
line indicates the exact value for infinite width, while the dotted
line includes the predicted finite size correction. It is nearly
indistinguishable from the best fit to a linear function (solid
line). (b) shows the values \protect$a_2(\width)$ for the true longest
common subsequence problem including the disorder correlations. Here,
the dashed and dotted lines show the upper and lower bounds for the
infinite system given by Dan{\v c}{\'\i}k~\protect\cite{danc94b} and
Boutet de Monvel~\protect\cite{demo99,demo00} respectively.}
\label{fig_finisizeplots}
\end{center}
\end{figure}

After we have convinced ourselves, that we can estimate the
$\widehat{a}_c(\width)$ to within a precision of $\pm10^{-6}$ for all
$\width$, we can study the dependence of $\widehat{a}_c(\width)$ on
the width $\width$ of the lattice.  Fig.~\ref{fig_finisizeplots}(a)
shows these numerically obtained values of $\widehat{a}_2(\width)$ for
an alphabet of size $c=2$ as a function of $1/\width$. The points
clearly fall on a straight line over the whole range of lattice widths
from $\width=128$ to $\width=4096$. A simple least square fit to a
straight line yields
\begin{equation}\label{eq_fittoa2hat}
\widehat{a}_2(\width)\approx
(0.8284270\pm0.0000009)-(0.0848\pm0.0002)\frac{1}{\width}.
\end{equation}
This has to be compared to the exact results
\begin{displaymath}
\frac{2}{\sqrt{2}\!+\!1}\approx0.82842712\quad\mbox{and}\quad
\frac{\sqrt{2}\!-\!1}{2(\sqrt{2}\!+\!1)}\approx0.08579.
\end{displaymath}
We conclude that the value for infinite system size is extremely well
reproduced by taking into account the $1/\width$ finite size
correction.  Even the prefactor of the finite size correction can be
extracted with a reasonably high precision, although the error of the
fit for the finite size coefficient as given in
Eq.~(\ref{eq_fittoa2hat}) underestimates the true error. This is to be
expected since the finite size coefficient itself is subject to
further finite size corrections which yield an additional {\em systematic}
error in its estimation.

\section{The Chv\'atal-Sankoff constant}

In order to get a good numerical estimate of the real
Chv\'atal-Sankoff constant, we perform the same simulations for the
subtly correlated disorder $\eta(r,t)$ generated from pairs of
randomly chosen sequences. In the presence of these disorder
correlations, the dynamics of the $\difference(r,t)$ at a given width
$\width$ can still be interpreted as a Markov process. However, the
applicable state space comprises at a given $t$ the $\difference(r,t)$
for $r\in\{0,\ldots,2\width-1\}$ {\em and} the last $\width$ letters
chosen in each of the two sequences. It is much larger than in the
first passage percolation version. Thus, although the dynamics still
converges exponentially to the stationary state, the initial phase can
be much longer than in the first passage percolation case and more
care has to be taken to minimize the systematic error of numerical
estimates introduced by this effect.

Again, we collect estimates from $1,000$ pairs of random sequences and
apply the recursion equations~(\ref{eq_ldiffboolean1})
and~(\ref{eq_ldiffboolean2}) $10^7$ times for each sequence pair.
Then, we generate the same $10$ estimates of $a_2(\width=4096)$ as in
Sec.~\ref{sec_firstpassage} for the largest width $\width=4096$ we
plan to use in our simulations. Fig.~\ref{fig_decorrel}(b) shows the
dependence of these estimates on the number $t_1$ of recursion steps
performed before starting the averaging. We note two important
differences to the first passage percolation case: (i) The estimate
with $t_1=0$ is significantly different from the estimates with
$t_1\ge10^6$ and (ii) the individual statistical fluctuations are
larger than in the first passage percolation case (note the difference
in scale between Figs.~\ref{fig_decorrel}(a) and~(b).) Both findings
are to be expected due to the slower approach to the stationary
state. Since the first estimate is different from the other nine we
conclude that the characteristic number of steps $t_0$ necessary to
approach the stationary state is smaller than but comparable to
$10^6$. This slow decorrelation property also renders the average over
many recursion steps less effective yielding the larger
statistical fluctuations which we observe. These statistical
fluctuations are of the order of $\pm3.5\cdot10^{-6}$. Since the
systematic deviation of the estimate for $t_1=10^6$ is already smaller
than that and since the characteristics number of steps $t_0$ is
smaller than $10^6$, we can be sure that the systematic deviations of
all points with $t_1\ge2\cdot10^6$ are less than $10^{-6}$ due to the
exponential decay of these deviations. Then, averaging in
Eq.~(\ref{eq_localadvance}) over all $t\in]2\cdot10^6,10^7]$ yields
estimates of $a_c(\width)$ with a statistical and systematic error of
$\pm10^{-6}$.

The results of this procedure for all finite system sizes $\width$ are
shown in Fig.~\ref{fig_finisizeplots}(b). In addition to the larger
statistical error bars of the individual points slight corrections to
the $1/\width$ behavior are visible for $\width<160$. Nevertheless, we
get a good fit to a straight line in the regime $\width\ge160$ with
the result
\begin{equation}
a_2(\width)\approx(0.812653\pm0.000004)-(0.0520\pm0.0012)\frac{1}{\width}.
\end{equation}
Varying the fitting range towards larger $\width$ changes the infinite
$\width$ value of the measured Chv\'atal-Sankoff constant only within
the quoted error bars. This result is significantly larger than the
results of Dan{\v c}{\'\i}k~\cite{danc94b} and of Boutet de
Monvel~\cite{demo99,demo00} obtained for $4$ lattices of size up to
$10^6\times10^6$ and $10,000$ lattices of size $10^4\times10^4$
respectively.  As shown in Fig.~\ref{fig_finisizeplots}(b) the value
of the finite width Chv\'atal-Sankoff constant already significantly
exceeds their estimates of the infinite system size value at moderate
$\width\approx256$.

In the same way we get high precision values for the Chv\'atal-Sankoff
constant for larger alphabet sizes $c$. However, the finite size
corrections for larger alphabet size are larger and fits to a linear
law only settle to within the statistical error estimate for
$W\ge352$. The fitted values are summarized in
Tab.~\ref{tab_csvalues}. While our method produces estimates with a
much higher precision than the traditional method due to the fact that
we know the form of the finite size dependence, all our estimates for
$c\ge4$ are consistent with Boutet de Monvel's within his variations
between different assumed finite size corrections~\cite{demo99}.  We
observe that the differences between the Chv\'atal-Sankoff constant
$a_c$ and its counterpart $\widehat{a}_c$ for the uncorrelated
disorder are significant but become smaller with increasing alphabet
size. This is to be expected intuitively since the correlations
between the $\eta(r,t)$ become weaker with increasing alphabet size
(if we know the letter in one sequence and the information that we
have a mismatch this fixes the letter in the other sequence only up to
$c-1$ possible choices.) This dependence on the alphabet size is also
consistent with the fact that it apparently becomes more difficult to
take the correlations into account numerically and to get good
numerical estimates of the $a_c$ as the alphabet size decreases. This
materializes in the discrepancies between our high precision results
for $a_2$ and earlier estimates of the Chv\'atal-Sankoff constant
which are not present for the larger alphabet sizes.

\begin{table}[htbp]
\begin{center}
\begin{tabular}{|c|l|l|l|}
\hline
\raisebox{1.5mm}{\strut}$c$&$a_c$&$b_c$&$\widehat{a}_c$\\
\hline
2&0.812653(4)&0.052(1)&0.828427\\
4&0.654361(2)&0.122(1)&0.666667\\
8&0.515143(4)&0.197(2)&0.522408\\
16&0.396316(2)&0.268(1)&0.4\\
\hline
\end{tabular}
\end{center}
\caption{Numerical values for the Chv\'atal-Sankoff constant
\protect$a_c$ and its finite size correction \protect$b_c$ for
different alphabet sizes \protect$c$. The errors in parentheses are
statistical errors on the last digit. For comparison the counterpart
\protect$\widehat{a}_c=2/(\sqrt{c}+1)$ of the Chv\'atal-Sankoff
constant for first-passage percolation is also given.}
\label{tab_csvalues}
\end{table}

\section{Conclusions}

We conclude that in order to get numerical values of the
Chv\'atal-Sankoff constant with a high precision a proper handling of
the finite size effects is absolutely necessary. We showed how
concepts from statistical physics can help to tackle this problem.  We
moreover presented a highly effective multi-spin coding scheme which
further increases the precision of the simulations. The numerical
results given will be valuable in guiding and evaluating theoretical
approaches to understand the small but noticeable effect of the subtle
disorder correlations which are responsible for the difference between
the Chv\'atal-Sankoff constant and its counterpart in the
first-passage percolation problem.

\section*{Acknowledgments}

The author gratefully acknowledges valuable discussions with
P.~Grassberger and T.~Hwa and the hospitality of Rockefeller
University, New York. This work has been partially supported by a
Hoch\-schul\-son\-der\-pro\-gramm~III fellowship of the DAAD and by
the NSF through Grants No. DMR-9971456 and DBI-9970199.

\appendix

\section{Change of lattice}\label{app_changelattice}

\begin{figure}
\begin{center}
\includegraphics[width=\columnwidth]{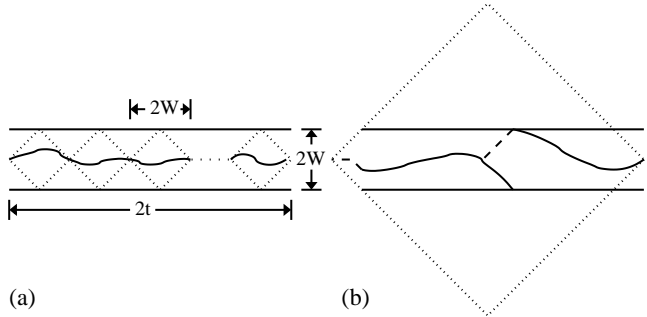}
\caption{Relation between optimal paths on the rectangular and the
diamond shaped lattice. (a) shows how a good path for the rectangular
lattice can be constructed by stringing together optimal paths on
diamond shaped lattices of size \protect$\width\times\width$. In (b) a
good path for a diamond shaped lattice of size
\protect$(t+\width)\times(t+\width)$ is constructed from the optimal
path on the rectangular lattice. The left end of the path is connected
to the tip of the diamond which can only increase the number of
matches. For each time the optimal path wraps around the periodic
boundary conditions the path has to be modified as shown by the
diagonal dashed line. For each such event, at most \protect$\width$
matches along the solid path can be missed.}
\label{fig_changelattice}
\end{center}
\end{figure}

In this appendix we will argue that the limiting quantity
\begin{equation}
a_c(\infty)=\lim_{\width\to\infty}a_c(\width)
=\lim_{\width\to\infty}\lim_{t\to\infty}
\frac{1}{t}\langle\ell(\width-1,2t)\rangle
\end{equation}
defined on the rectangular lattice shown in Fig.~\ref{fig_alignl}(b)
equals the Chv\'atal-Sankoff constant
\begin{equation}
a_c\equiv\lim_{N\to\infty}\frac{\langle L(N)\rangle}{N}
\end{equation}
defined on the diamond shaped lattice shown in
Fig.~\ref{fig_alignl}(a). To this end, we first consider a path from
the left edge of the rectangular lattice to the point $(\width-1,2t)$
as indicated in Fig.~\ref{fig_changelattice}(a) which is the optimal
path in each of the diamonds. Clearly, $\ell(\width-1,2t)$ can only be
larger than the number of matches along this specific path which
implies
\begin{equation}
\langle\ell(\width-1,2t)\rangle\ge\frac{t}{\width}\langle L(\width)\rangle
\end{equation}
or after dividing both sides by $t$ and taking the limit of $t\to\infty$
\begin{equation}
a_c(\width)\ge\frac{\langle L(\width)\rangle}{\width}.
\end{equation}
Taking the limit $\width\to\infty$ of this inequality yields
\begin{equation}\label{eq_acinftygeac}
a_c(\infty)\ge a_c.
\end{equation}

On the other hand, let us consider the best path from the left edge of
the rectangular lattice to the point $(\width-1,2t)$, i.e., the path
with $\ell(\width-1,2t)$ matches. As shown in
Fig.~\ref{fig_changelattice}(b), we can construct a path which
connects the two end points in a diamond shaped lattice of size
$(t+\width)\times(t+\width)$ from this path. Since the length of the
longest common subsequence of the two associated sequences of length
$t+\width$ can only be longer than the common subsequence corresponding
to this path we get
\begin{equation}
\langle L(t+\width)\rangle\ge\langle\ell(\width-1,2t)\rangle-
\width\langle n_{\mathrm{wrap}}(t)\rangle,
\end{equation}
where $n_{\mathrm{wrap}}(t)$ is the number of times that the best path
from the left edge to $(\width-1,2t)$ ``wraps around'' the periodic
boundary conditions. Dividing by $t$ yields
\begin{equation}
\frac{t+\width}{t}\frac{\langle L(t+\width)\rangle}{t+\width}\ge
\frac{\langle\ell(\width-1,2t)\rangle}{t}-
W\frac{\langle n_{\mathrm{wrap}}(t)\rangle}{t}
\end{equation}
which in the limit $t\to\infty$ leads to
\begin{equation}\label{eq_withwrapping}
a_c\ge a_c(W)-
W\lim_{t\to\infty}\frac{\langle n_{\mathrm{wrap}}(t)\rangle}{t}.
\end{equation}
From the analogy with the directed polymer in a random medium it is
well known that the displacement of a typical optimal path of length
$t$ scales like $t^{2/3}$~\cite{hwa96,krug91a,dras97,dras98}. Thus, we
expect one wrapping around every $\width^{3/2}$ steps along the
lattice, i.e.,
\begin{equation}
\langle n_{\mathrm{wrap}}(t)\rangle\sim\width^{-3/2}t.
\end{equation}
Therefore, if we take the limit $\width\to\infty$ in
Eq.~(\ref{eq_withwrapping}) the second term on the right hand side
vanishes and we are left with
\begin{equation}
a_c\ge a_c(\infty).
\end{equation}
Together with Eq.~(\ref{eq_acinftygeac}) this establishes the equality
between the infinite width limit of the $a_c(\width)$ and the
Chv\'a\-tal-Sankoff constant $a_c$.

\end{document}